\documentclass[a4paper,11pt]{article}
\usepackage{pos}

\title{Ab-initio study of dibaryons with highest bottom number \note{MITP-22-086, TIFR/TH/22-42}}

\author*[a]{M.\ Padmanath}
\author[b]{Nilmani Mathur}
\author[b]{Debsubhra Chakraborty}

\affiliation[a]{Helmholtz Institut Mainz, Staudingerweg 18, 55128 Mainz, Germany, \\
GSI Helmholtzzentrum für Schwerionenforschung, PlanckStr. 1, Darmstadt (Germany)}

\affiliation[b]{Tata Institute of Fundamental Research, Homi Bhabha Road, Colaba,  \\ 
Mumbai 400005, India.}

\emailAdd{pmadanag@uni-mainz.de}

\abstract{We present the first lattice study of dibaryons with highest bottom number. Utilizing 
a set of state-of-the-art lattice QCD ensembles and methodologies, we determine the ground state of 
dibaryon composed of two $\Omega_{bbb}$ baryons. We extract the related scattering amplitude 
in the $^1S_0$ channel and find a sub-threshold pole, which signifies an unambiguous evidence for 
a deeply bound $\Omega_{bbb}-\Omega_{bbb}$ dibaryon. The binding energy of such a state as dictated 
by this pole singularity is found to be -81($^{+14}_{-16}$) MeV. We quantify various systematic 
uncertainties involved in this determination, including those related to the excited state 
contamination and Coulomb repulsion between the bottom quarks.}

\FullConference{%
 The 39th International Symposium on Lattice Field Theory, LATTICE2022 \\
  08$^{th}$-13$^{th}$ August, 2022, Bonn, Germany
}

\newcommand\bef{\begin{figure}}
\newcommand\eef[1]{\label{fg:#1}\end{figure}}
\newcommand\bec{\begin{center}}
\newcommand\eec{\end{center}}
\newcommand\besf{\begin{subfigure}}
\newcommand\eesf[1]{\label{sfg:#1}\end{subfigure}}
\newcommand\beq{\begin{equation}}
\newcommand\eeq[1]{\label{#1}\end{equation}}
\newcommand\beqa{\begin{eqnarray}}
\newcommand\eeqa[1]{\label{#1}\end{eqnarray}}
\newcommand\bet{\begin{table}}
\newcommand\eet[1]{\label{tb:#1}\end{table}}
\newcommand\best{\begin{subtable}}
\newcommand\eest[1]{\label{stb:#1}\end{subtable}}
\newcommand\betb{\begin{center}\begin{tabular}}
\newcommand\eetb{\end{tabular}\end{center}}
\newcommand\beit{\begin{itemize}}
\newcommand\eeit{\end{itemize}}

\newcommand\fgn[1]{Figure \ref{fg:#1}}
\newcommand\eqn[1]{eq.\ (\ref{#1})}

\newcommand\tbn[1]{Table \ref{tb:#1}}

\newcommand{\cref}[1]{\bec  \textcolor{black}{\small #1}\eec}

\definecolor{DarkGreen}{rgb}{0.00,0.29,0.00}
\definecolor{DarkRedfooter}{rgb}{0.60,0.00,0.00}
\definecolor{DarkRed}{rgb}{0.60,0.00,0.00}
\definecolor{DarkRedtitle}{rgb}{0.85,0.00,0.00}

\def\prsp#1#2%
  {\mathop{}%
   \mathopen{\vphantom{#2}}^{#1}%
   \kern-\scriptspace%
   #2}

\begin{document}
\maketitle

\section{Introduction}

Understanding baryon-baryon interactions from first-principles is of prime interest in nuclear and
astrophysics, as they form the foundation towards a fundamental explanation of why atomic nuclei exist. 
Simplest systems in which such interactions can be studied transparently are dibaryons, which have 
a baryon number 2. Despite decades of experimental efforts, only one bound dibaryon has been established
(Deuteron). Even so, there might be other bound dibaryons, particularly in the heavy sector, that exist 
in Nature and are yet to be discovered.

As of today, there are a handful of lattice QCD investigations of dibaryons in the light and the strange
sector, mostly performed at unphysically heavy quark masses, covering channels such as Deuteron, 
dineutron, H-dibaryon, etc. {\it c.f.} Ref \cite{Green:2021qol}. An important lesson learned from 
these studies is that discretization effects could be substantial in these systems, which when 
unaddressed can lead to large discrepancies. Among the heavy baryons, $\Omega_{ccc}$ and $\Omega_{bbb}$
serves a promising domain to study the nonperturbative features of QCD, free of the light quark dynamics.
On the same ground, a system of two $\Omega_{ccc}$ and two $\Omega_{bbb}$ baryons are interesting as 
they serve as useful platforms in understanding the baryon-baryon interactions in chiral dynamics free
environment. Such a system in the light sector is more complicated, since $\Delta$ is a resonance and 
that there are elastic threshold corresponds to four particle final states. Scattering of $\Omega_{sss}$
baryons has been addressed in a few lattice QCD works \cite{Buchoff:2012ja,HALQCD:2015qmg,Gongyo:2017fjb}
following different methodologies. Although they arrive at conflicting conclusions on whether 
the interaction between $\Omega_{sss}$ baryons is weakly attractive or weakly repulsive, a common
consensus from these studies is that the interactions are very weak\footnote{The word `weak' used
throughout this article refers to the strength of the interaction, and has no connection to the real 
weak interaction in the Standard Model of Particle Physics.} in Nature. Another recent lattice study of
$\Omega_{ccc}$ baryons scattering reported a very shallow bound state in the $^1S_0$ channel
\cite{Lyu:2021qsh}. While all these investigations suggest weak interactions in systems with quark 
masses ranging from light to charm, several lattice studies in the recent years on heavy dibaryons
\cite{Junnarkar:2019equ,Junnarkar:2022yak} and heavy tetraquarks \cite{Francis:2016hui,Junnarkar:2018twb}
have shown that multi-hadron systems with multiple bottom quarks can have deep binding. Naturally, 
an investigation of interactions between $\Omega_{bbb}$ baryons would be very timely, as this can shed
light on the behaviour of such interactions in the heavy sector.  

In this work, we perform a lattice QCD calculation of scalar dibaryons with highest number of 
bottom quarks. We name such a dibaryon system as $\mathcal{D}_{6b}$. The elastic threshold 
corresponds to the $\Omega_{bbb}-\Omega_{bbb}$ channel. 

\section{Methodology}\label{method}

\underline{\textit{Ensembles}}:
We utilize four ensembles with dynamical u/d, s, and c quark fields generated by the MILC collaboration, 
with HISQ fermion action \cite{Bazavov:2012xda}. The specifics of the lattice ensembles are as shown in 
\fgn{ensmap}.
\bef
\centering
\includegraphics[scale=0.35]{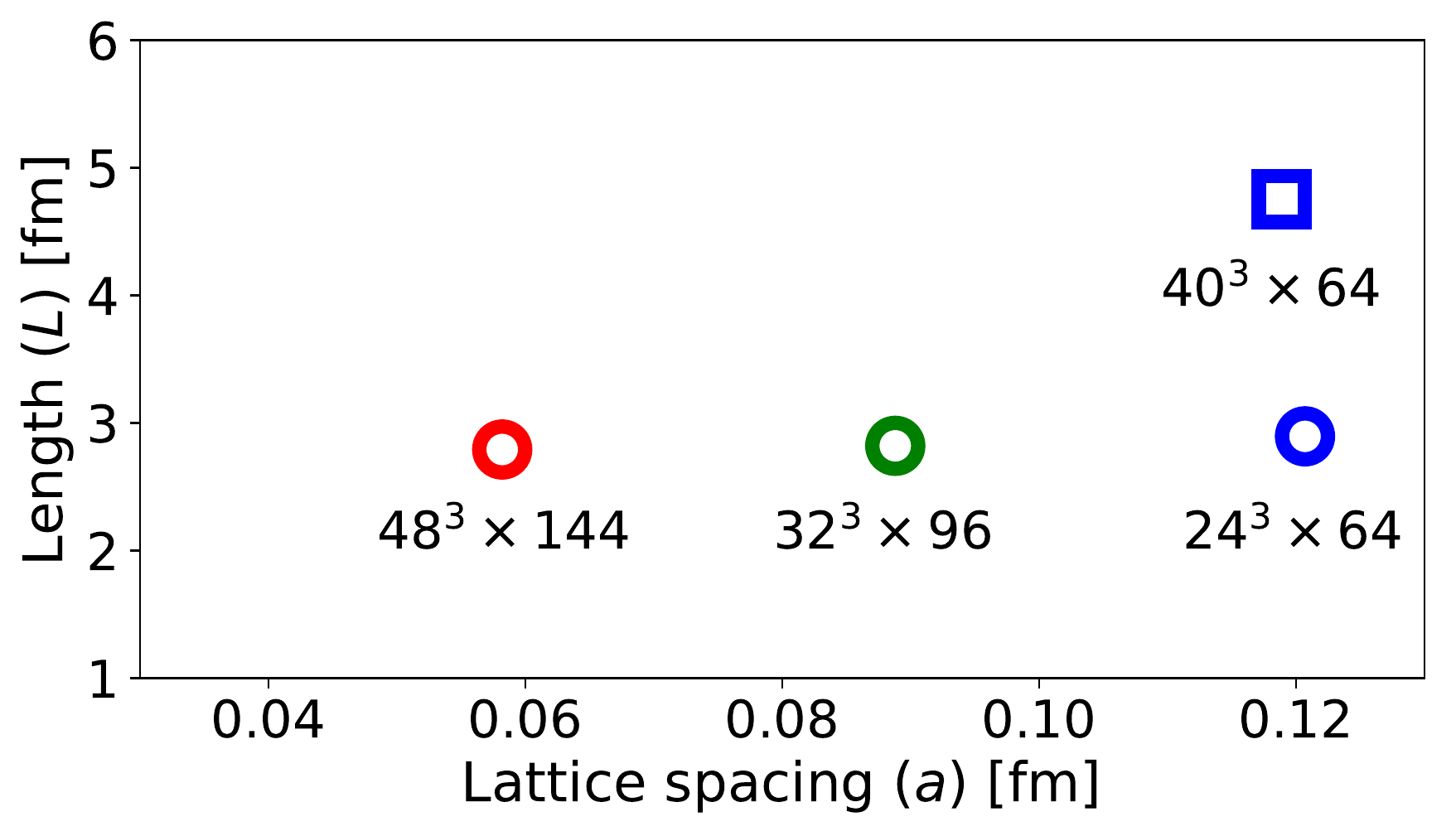}
\caption{Lattice QCD ensembles utilized in this work presented in the $L$ (spatial extent) versus $a$ 
(lattice spacing) landscape.}
\eef{ensmap}

\underline{\textit{Quark propagators}}: Quark propagators were computed using the time evolution of 
an NRQCD Hamiltonian, including improvements up to $\mathcal{O}(\alpha_sv^4)$ on Coulomb gauge fixed 
wall sources and multiple source timeslices. The bare quark mass was tuned using the spin averaged 
kinetic mass of 1S bottomonia states and was found to reproduce the 1S bottomonia hyperfine splitting accurately.

\underline{\textit{Two point correlation functions}}:
Hadron spectroscopy programs in lattice QCD such as ours proceed through the computation of two point 
correlation functions of the form 
\beq
C_{\mathcal{O}}(t_f-t_i) = \sum_{\vec{x}_f} e^{-i\vec{p}.\vec{x}_f}\langle 0 | 
\mathcal{O}(\vec{x}_f,t_f)\bar{\mathcal{O}}(t_i)|0 \rangle,
\eeq{Eq:Domega2}
where $\mathcal{O}$ are the interpolating operators with the desired quantum numbers. 
At the sink time slice, we utilize several different quark field smearing procedures to identify 
the reliable ground state plateau and quantify any possible excited state contamination (see 
Ref.~\cite{Mathur:2022nez} for more details). The single and the dibaryon ground state energies in 
the finite-volume are obtained by fitting the averages of correlation function with a single exponential 
at large times.

\underline{\textit{Interpolating operators}}:
A nondisplaced nonrelativistic operator with $J^P=3/2^+$ \cite{Buchoff:2012ja} was utilized for $\Omega_{bbb}$ baryon. 
The color space is trivially antisymmetric for a baryon. The flavor and the spatial structure of 
the operator is trivially symmetric. The remaining spin space for a nonrelativistic operator allows 
only J=3/2 spin, which is symmetric as shown below. 
\bet[h]
\centering
\begin{tabular}{@{\hspace{0.3cm}}c | @{\hspace{0.3cm}}c |@{\hspace{0.3cm}}c }\hline \hline
  $\mathcal{O}_{\Omega_{bbb}}$ & $|s_1s_2s_3\rangle$ & $|J, J_z\rangle$ \\\hline
  $\chi_1$ & $|111\rangle$   & $|3/2,+3/2\rangle$ \\
  $\chi_2$ & $|112\rangle_S$ & $|3/2,+1/2\rangle$ \\
  $\chi_3$ & $|122\rangle_S$ & $|3/2,-1/2\rangle$ \\
  $\chi_4$ & $|222\rangle$   & $|3/2,-3/2\rangle$ \\
  \hline
\end{tabular}
\caption{Four rows of the $\mathcal{O}_{\Omega_{bbb}}$ baryon operator in the $^1H^+$ finite-volume irrep. 
$\chi_i$ refers to different azimuthal components and $s_i$ refer to the spin components of 
the $i^{th}$ quark. For NRQCD action, only the upper two components of the Dirac spinor are nonzero 
and are referred using 1 and 2 respectively. All spin components of the total operator is symmetrized, 
and are highlighted with a subscript $S$. The third column shows the total spin and the azimuthal 
component of each rows of the operator.}
\eet{Obbb_op}

Assuming dominance of s-wave scattering in the $\Omega_{bbb}$ interactions near the two baryon threshold, 
we build the two hadron operator relevant for scalar $\mathcal{D}_{6b}$ channel using simple spin 
algebra as follows. 
\beq
\mathcal{O}_{\mathcal{D}_{6b}^{J=0}} = 
\frac12~\Bigl[\chi_{\frac32}\chi_{-\frac{3}2}+\chi_{-\frac12}\chi_{\frac12}-\chi_{\frac12}\chi_{-\frac12}-
\chi_{-\frac32}\chi_{\frac{3}2}\Bigr].
\eeq{Eq:Domega12}

\section{Results}\label{results}

\underline{\textit{Signal quality and fit estimates}}:
We present the signal quality and the ground state saturation in baryon and the dibaryon correlation
functions in terms of the effective mass, defined as $am_{eff}=log[C(\tau)/C(\tau+1)]$, in \fgn{effmass}. 
Here $C(\tau)$ is the correlation function at source sink separation time $\tau$. Signal saturation and 
a clear energy gap between the noninteracting two-baryon level and the interacting dibaryon system in 
the finest ensemble is evident. This pattern is observed on all four ensembles we study, which
unambiguously point to a finite volume energy level in the interacting dibaryon system below 
the threshold.  In order to ensure the identification of the real ground state plateau, we perform 
the same analysis using different smearing procedures (see Ref [11] for details). This helps us 
estimating the errors related to possible excited state contamination arising from incorrect
identification of the ground state plateau.

\bef[h]
\centering
\includegraphics[scale=0.35]{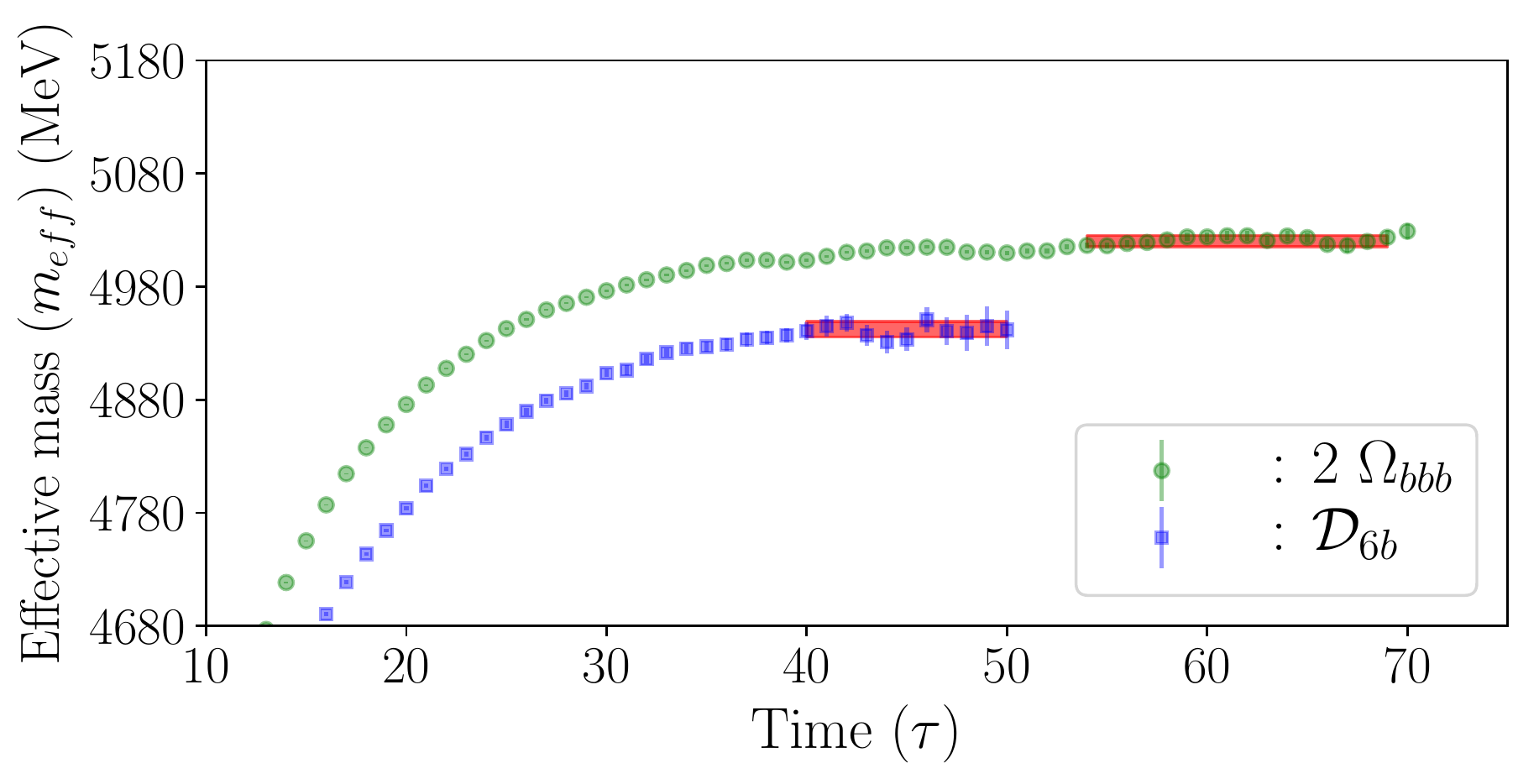}
\vspace*{-0.09in}
\caption{Effective masses of the non-interacting two-baryon (green) and dibaryon (blue) ground states 
in the finite volume on the finest lattice ensemble. A negative energy shift in the interacting case 
is clear, indicating attractive interations. The solid bands present the fit results and fitting windows.}
\eef{effmass}

Single exponential fits to $C(\tau)$ yield the mass estimates in lattice units for the baryon and dibaryon
ground states. The use of NRQCD formulation for quark fields implies an additive normalization 
proportional to the number of heavy quarks within the hadron realized. To this end, we determine the 
energy splittings between the baryon and dibaryon ground state mass estimates. Note that by taking such 
energy differences, it automatically removes of the additive normalization in the energy estimates 
intrinsic to the NRQCD formulation. This energy splitting $\Delta E=M_{\mathcal{D}_{6b}} - 2 
M_{\Omega_{bbb}}$ in physical units as determined on different lattice are tabulated in \tbn{delE}. 

\bet[h]
\centering
\begin{tabular}{@{\hspace{0.3cm}}c @{\hspace{0.3cm}}|@{\hspace{0.3cm}} c @{\hspace{0.7cm}}|| @{\hspace{0.7cm}}c @{\hspace{0.3cm}}| @{\hspace{0.3cm}}c @{\hspace{0.3cm}}}\hline \hline
  Ensemble & $\Delta E$ & Ensemble & $\Delta E$\\
  \hline \hline
  $24^3 \times 64$ & $ -61(11)$ & $40^3 \times 64$ & $ -62(7)$ \\
  $32^3 \times 96$ & $ -68(9)$  & $48^3 \times 144$ & $ -71(7)$\\
  \hline
\end{tabular}
\caption{Energy splittings $\Delta E = M_{\mathcal{D}_{6b}} - 2M_{\Omega_{bbb}}$
in MeV on different ensembles.}
\eet{delE}

\underline{\textit{Scattering analysis}}: 
The existence and the properties of a hadron from finite volume spectrum is determined in terms of 
pole singularities in the relevant scattering amplitudes across the complex Mandelstam $s$-plane. 
To this end, we follow L\"uscher’s finite-volume formalism. The s-wave scattering amplitude is given 
by $t=(\cot\delta_0-i)^{-1}$, and a pole in $t$ related to a bound state happens when $k.\cot\delta_0 = 
-\sqrt{-k^2}$. Here $\delta_0(k)$ is the two spin $\Omega_{bbb}$ scattering phase shift leading to 
$J^P=0^+$ is related to the finite-volume energy eigenvalues as 
\beq
k~cot(\delta_0(k)) = \frac{2Z_{00}(1;(\frac{kL}{2\pi})^2))}{L\sqrt{\pi}}.
\eeq{luscher}
Here $k$ is the momentum of the $\Omega_{bbb}$ baryon in the centre of momentum frame and 
is given by $k^2 = \frac{\Delta E}{4} (\Delta E+4M_{\Omega_{bbb}}^{phys})$, and $M_{\Omega_{bbb}}^{phys}$
is the mass of $\Omega_{bbb}$ baryon in the continuum limit determined independently from the 
respective correlation functions. $Z$s are the L\"uscher's zeta functions \cite{Luscher:1990ux}. 

We parameterize $k~cot\delta_0$ either as a constant or as a constant plus a linear term in the 
lattice spacing to determine possible cutoff systematics in our finite volume treatment. We perform 
several different fits involving different subsets of the four energy splittings listed earlier, with 
the two different fits forms. All of the fits indicate the existence of a deeply bound state pole in 
this channel. The best fit is found to be the one that incorporates all the energy splittings, and
incorporates the lattice spacing dependence of the scattering length $a_0$ with a linear 
parametrization $k~cot\delta_0 = -1/a_0^{[0]} - a/a_0^{[1]}$. The fit quality turns out to be 
$\chi^2/d.o.f= 0.7/2$ and the estimates in this case and the resultant binding energy are
\beqa
a_0^{[0]} = 0.18(^{+0.02}_{-0.02})~\text{fm}, &\quad&  a_0^{[1]}
=-0.18(^{+0.18}_{-0.11})~\text{fm}^2,\label{scatlen} \nonumber \\
\text{and~~} \Delta E_{\mathcal{D}_{6b}} &=& -81(^{+14}_{-16})~\text{MeV}.
\eeqa{beDO}
We present the details of our main results in the below figure. 
\bef[h!]
\centering
\includegraphics[height=7.5cm,width=8.8cm]{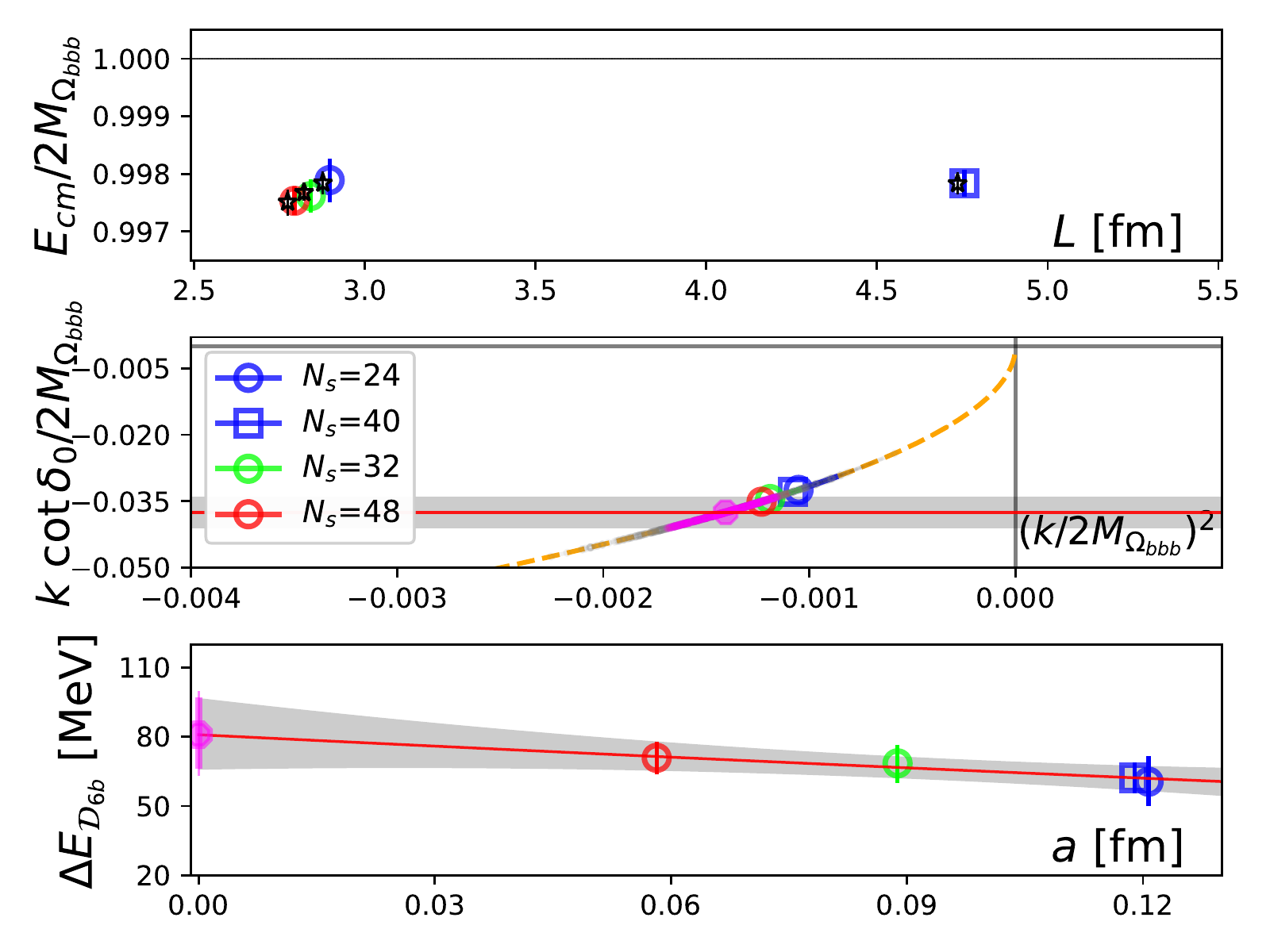}
\caption{Results from the finite-volume scattering analysis. Top: Comparison of the simulated energy levels
(large symbols) with the energy levels (black stars) analytically reconstructed using Eq. (\ref{scatlen}), 
indicating the quality of the scattering analysis fit. Middle: $k~cot\delta_0$ versus $k^2$ in units of
energy of the threshold ($2M_{\Omega_{bbb}}$) and information on poles in $t$ indicated by magenta 
symbols. Bottom: Continuum extrapolation of the binding energy in \eqn{beDO} determined from fitted
scattering amplitude in Eq. (\ref{scatlen}).}
\eef{pcot1}

\underline{\textit{Various systematics}}: Now we discuss various systematic uncertainties related to 
this calculation. We use state-of-the-art lattice QCD ensembles in this study, together with an NRQCD 
Hamiltonian with improvement coefficients up to $\mathcal{O}(\alpha_sv^4)$ for the time evolution of 
the bottom quark fields. This setup has been demonstrated to reproduce the bottomonium hyperfine
splittings correctly with an uncertainty of about 6 MeV \cite{Mathur:2022nez}. A faithful reproduction 
of the hyperfine splitting reflects small and controllable discretization effects. Note that the energy
splittings that we compute and utilize in our scattering analysis have reduced systematics. For 
the heavy dibaryons, statistical errors, fit-window errors and possible excited state contamination in 
the identified energy plateau are the main sources of error. We arrive at robust identification of 
the ground state plateau by studying the ground state signals for the same state with different smearing
programs. The systematics arising from excited state effects are gauged from the variation in 
the continuum limit estimates determined from different smearing and analysis procedures utilized.

This dibaryon being extremely heavy, bound and have an electric charge -2, effects from Coulombic 
repulsion could be substantial. We investigate possible corrections due to this following an analysis 
procedure as in Ref. \cite{Lyu:2021qsh}. The interaction between the $\Omega_{bbb}$ baryons is modelled 
with an attractive multi-Gaussian potential $V_s$ with its parameters tuned to reproduce the quantum
mechanical bound state with binding energy $-81(_{-16}^{+14})$ MeV. The Coulombic interaction is further
modelled as in Ref. \cite{Lyu:2021qsh}, and its effects are determined by studying the ground state
solutions of the total potential. In \fgn{coul1}, we show the parameterized potentials $V_s$ and Coulombic
interaction and their combination, together with the radial probabilities of the ground state 
wave-functions in $V_s$ and combined potentials. It is evident that the Coulombic interaction hardly 
affects $\Omega_{bbb}$ baryon interaction potential $V_s$ and its ground state radial probability
densities. We find that the deviations in binding energy with inclusion of such Coulombic interaction are
found to be between $5-10$ MeV. Other systematic uncertainties related to the continuum extrapolation 
fit forms, scale setting, quark mass tuning and electromagnetic effects are found to be less than or equal
to 12 MeV.

\bef[t]
\centering
\includegraphics[scale=0.35]{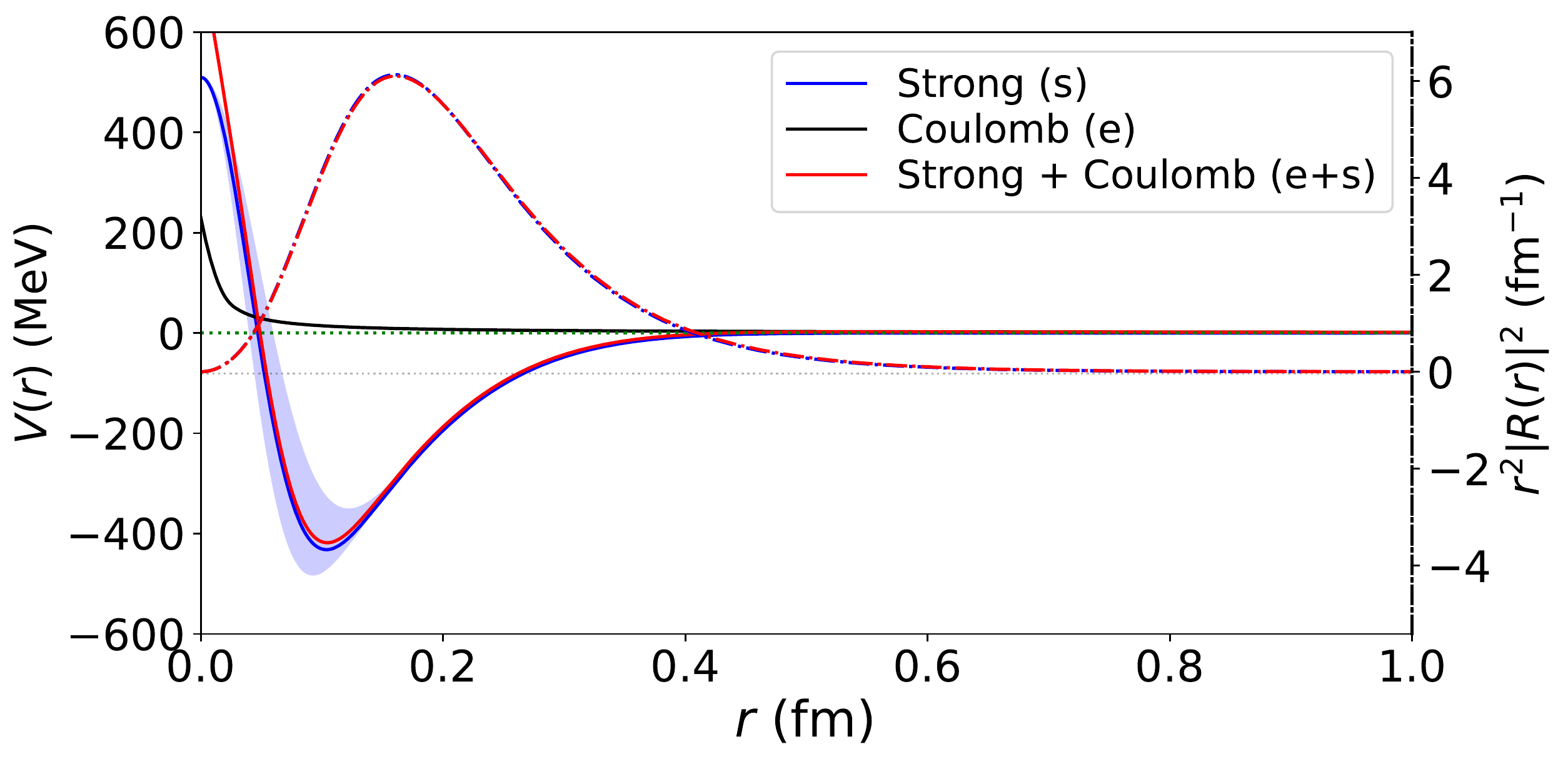}
\caption{ Coulombic potential ($V_e$), the parameterized $\Omega_{bbb}$ baryon interaction
potentials ($V_s$) and their sum are shown by the solid black, blue and red curves, respectively. 
The shaded band is the variation of $V_s$ with respect to its parameters. The radial probability densities 
of the quantum mechanical ground state wave-functions of $V_s$ and combined potentials are represented by
the dotdashed curves.}
\eef{coul1}

\section{Summary and conclusions} \label{summary}
We present a first investigation of interactions between $\Omega_{bbb}$ baryons. We find a deeply bound
dibaryon $\mathcal{D}_{6b}$ in the $^1S_0$ channel. The relevant scattering amplitude is extracted
following the L\"uscher's formalism which features a bound state pole with binding energy 
$-81(^{+14}_{-16})(14)$ MeV. Complementary measurement and analysis procedures are utilized in identifying
the real ground state plateau to ensure the robustness of our results. All possible systematics, 
including but not limited to fitting-window errors and excited state effects are studied. The resultant
uncertainties added in quadrature are quantified in the second parenthesis. Unlike the light dibaryons
with a handful of lattice calculations with conflicting observations, our conclusion on the existence of 
a deeply bound state $\mathcal{D}_{6b}$ would be robust to calculations such as those using a variational
approach. Considering the observation of weakly interacting nature in equivalent systems
($\mathcal{D}_{6s}$ and $\mathcal{D}_{6c}$) \cite{Buchoff:2012ja,HALQCD:2015qmg,Gongyo:2017fjb}, future
studies of quark mass dependence of this system would be very appealing and could shed further light into
the change in quark dynamics at different energy scales.

\acknowledgments
I would like to thank the organizers of this conference for a very enjoyable conference,
and the participants for many illuminating discussions, in particular S.~Collins, J.~Green, 
F.~K.~Guo(online), L.~Leskovec, L.~Liu(online), D.~Mohler, S.~Prelovsek, and M.~Pflaumer 
for discussions. This work is supported by the Department of Atomic Energy, Government of India, 
under Project Identification Number RTI 4002. Computations were carried out on the Cray-XC30 of 
ILGTI, TIFR. 
\bibliographystyle{JHEP}
\bibliography{lattice2022}

\providecommand{\href}[2]{#2}\begingroup\raggedright\begin{thebibliography}{10}

\bibitem{Green:2021qol}
J.~R. Green, A.~D. Hanlon, P.~M. Junnarkar and H.~Wittig,
  \href{https://doi.org/10.1103/PhysRevLett.127.242003}{\emph{Phys. Rev. Lett.}
  {\bfseries 127} (2021) 242003}
  [\href{https://arxiv.org/abs/2103.01054}{{\ttfamily 2103.01054}}].

\bibitem{Buchoff:2012ja}
M.~I. Buchoff, T.~C. Luu and J.~Wasem,
  \href{https://doi.org/10.1103/PhysRevD.85.094511}{\emph{Phys. Rev. D}
  {\bfseries 85} (2012) 094511}
  [\href{https://arxiv.org/abs/1201.3596}{{\ttfamily 1201.3596}}].

\bibitem{HALQCD:2015qmg}
{\scshape HAL QCD} collaboration, M.~Yamada, K.~Sasaki, S.~Aoki, T.~Doi,
  T.~Hatsuda, Y.~Ikeda et~al., ,
  \href{https://doi.org/10.1093/ptep/ptv091}{\emph{PTEP} {\bfseries 2015}
  (2015) 071B01} [\href{https://arxiv.org/abs/1503.03189}{{\ttfamily
  1503.03189}}].

\bibitem{Gongyo:2017fjb}
{\scshape HAL QCD Collaboration} collaboration, S.~Gongyo, K.~Sasaki, S.~Aoki,
  T.~Doi, T.~Hatsuda, Y.~Ikeda et~al., ,
  \href{https://doi.org/10.1103/PhysRevLett.120.212001}{\emph{Phys. Rev. Lett.}
  {\bfseries 120} (2018) 212001}.

\bibitem{Lyu:2021qsh}
Y.~Lyu, H.~Tong, T.~Sugiura, S.~Aoki, T.~Doi, T.~Hatsuda et~al.,
  \href{https://doi.org/10.1103/PhysRevLett.127.072003}{\emph{Phys. Rev. Lett.}
  {\bfseries 127} (2021) 072003}
  [\href{https://arxiv.org/abs/2102.00181}{{\ttfamily 2102.00181}}].

\bibitem{Junnarkar:2019equ}
P.~Junnarkar and N.~Mathur,
  \href{https://doi.org/10.1103/PhysRevLett.123.162003}{\emph{Phys. Rev. Lett.}
  {\bfseries 123} (2019) 162003}
  [\href{https://arxiv.org/abs/1906.06054}{{\ttfamily 1906.06054}}].

\bibitem{Junnarkar:2022yak}
P.~M. Junnarkar and N.~Mathur,
  \href{https://doi.org/10.1103/PhysRevD.106.054511}{\emph{Phys. Rev. D}
  {\bfseries 106} (2022) 054511}
  [\href{https://arxiv.org/abs/2206.02942}{{\ttfamily 2206.02942}}].

\bibitem{Francis:2016hui}
A.~Francis, R.~J. Hudspith, R.~Lewis and K.~Maltman,
  \href{https://doi.org/10.1103/PhysRevLett.118.142001}{\emph{Phys. Rev. Lett.}
  {\bfseries 118} (2017) 142001}
  [\href{https://arxiv.org/abs/1607.05214}{{\ttfamily 1607.05214}}].

\bibitem{Junnarkar:2018twb}
P.~Junnarkar, N.~Mathur and M.~Padmanath,
  \href{https://doi.org/10.1103/PhysRevD.99.034507}{\emph{Phys. Rev.}
  {\bfseries D99} (2019) 034507}
  [\href{https://arxiv.org/abs/1810.12285}{{\ttfamily 1810.12285}}].

\bibitem{Bazavov:2012xda}
{\scshape MILC Collaboration} collaboration, A.~Bazavov, C.~Bernard,
  J.~Komijani, C.~DeTar, L.~Levkova, W.~Freeman et~al., ,
  \href{https://doi.org/10.1103/PhysRevD.87.054505}{\emph{Phys. Rev. D}
  {\bfseries 87} (2013) 054505}.

\bibitem{Mathur:2022nez}
N.~Mathur, M.~Padmanath and D.~Chakraborty,
  \href{https://arxiv.org/abs/2205.02862}{{\ttfamily 2205.02862}}.

\bibitem{Luscher:1990ux}
M.~Luscher, \href{https://doi.org/10.1016/0550-3213(91)90366-6}{\emph{Nucl.
  Phys. B} {\bfseries 354} (1991) 531}.

\end{thebibliography}\endgroup

\end{document}